\documentclass[conference]{IEEEtran}
\IEEEoverridecommandlockouts
\usepackage{cite}

\usepackage{amsmath}
\usepackage{comment}
\usepackage{graphicx}
\usepackage{url}
\def\x{{\mathbf x}}

\def\C{{\mathcal C}}
\def\L{{\cal L}}

\def\h{\mathbf h}

\def\Z{\mathcal{Z}}

\def\Z{Z}

\def\CTC{\mathrm{CTC}}
\def\ctc{\mathrm{ctc}}

\def\ctc{\mathrm{ctc}}

\def\enc{\mathrm{enc}}

\def\BibTeX{{\rm B\kern-.05em{\sc i\kern-.025em b}\kern-.08em
    T\kern-.1667em\lower.7ex\hbox{E}\kern-.125emX}}
\begin{document}

\title{Spiralformer: Low Latency Encoder for \\Streaming Speech Recognition with \\Circular Layer Skipping and Early Exiting}

\author{\IEEEauthorblockN{Emiru Tsunoo, Hayato Futami, Yosuke Kashiwagi}
\IEEEauthorblockA{\textit{Sony Group Corporation}, Japan}
\and
\IEEEauthorblockN{Siddhant Arora, Shinji Watanabe}
\IEEEauthorblockA{\textit{Carnegie Mellon University}, USA}
}

\maketitle
\begin{abstract}
For streaming speech recognition, a Transformer-based encoder has been widely used with block processing. Although many studies addressed improving emission latency of transducers, little work has been explored for improving encoding latency of the block processing. We seek to reduce latency by frequently emitting a chunk with a small shift rather than scarce large-chunk emissions, resulting in higher computational costs. To efficiently compute with the small chunk shift, we propose a new encoder, Spiralformer, tailored for block processing by combining layer dropping and early exiting. We skip layer computation in a cyclic manner and shift the computed layer in each block spirally, which completes computation for all the layers over the block processing. Experimentally, we observed that our method achieved 21.6\% reduction in the averaged token emission delay in Librispeech, and 7.0\% in CSJ, compared with the baseline with similar computational cost and word error rates.
\end{abstract}

\section{Introduction}
Streaming automatic speech recognition (ASR) is required in many real-world use cases, such as real-time transcription of broadcast content.
Many streaming end-to-end approaches adopt blockwise processing for encoders \cite{povey18,dong19, tsunoo19, miao2020,shi2021emformer}.
The encoders are then combined with connectionist temporal classification (CTC) \cite{graves06, miao15, amodei16, arora2023semi}, or transducers \cite{graves13rnnt, rao17,zhang2020transformer}, which have an affinity for streaming processing with frame-wise computation. 
Because the transducer particularly introduces delays in token emission timing, some studies regularize the model to reduce emission latency \cite{li2020towards,sainath2020emitting,yu2021fastemit,song2023trimtail}.
However, less work has been done to reduce latency in the blockwise processing encoders.

\begin{figure}[t]
  \centering
  \includegraphics[width=1.0\linewidth]{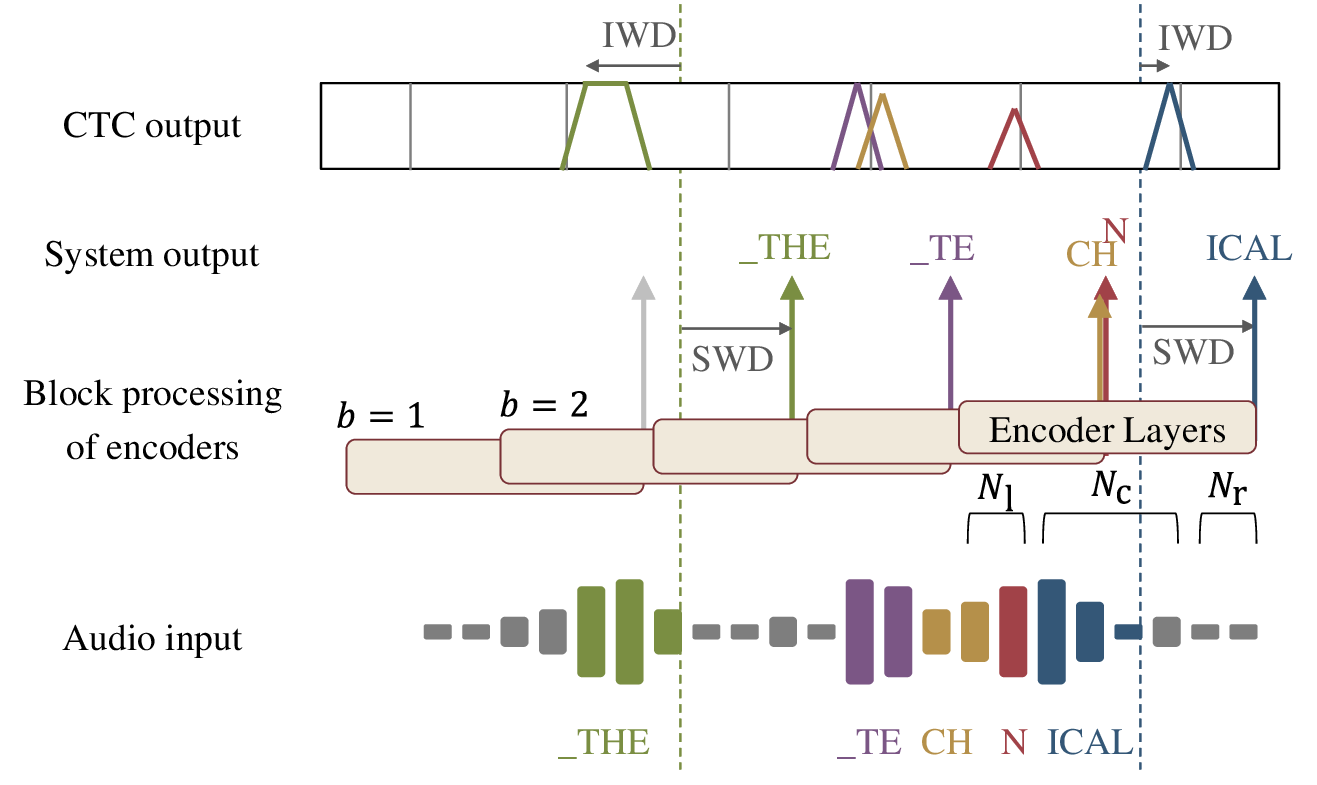}
  \caption{Delays in blockwise processing. $N_l$ is the number of left context, $N_c$ is the number of non-overlapping chunks, and $N_r$ is the number of right context (look-ahead frames). Internal Word emission Delay (IWD) is a delay of internal modeling, and System Word emission Delays (SWD) is an overall delay of the entire system. Although CTC emissions are aligned with the audio input, the block processing hinders the emission at most the shift size and right context ($N_c+N_r$).}
  \label{fig:delay}
\vspace{-0.3cm}
\end{figure}

In the competition of word error rates (WERs) on ASR test sets, the encoders typically use a similar choice of design for block processing to minimize the WERs, which often sacrifice latency.
An example of block processing is illustrated in Fig.~\ref{fig:delay}.
In each block processing, the encoder consumes $N_c$-frame non-overlapping chunks, and also takes into account overlapping left context of $N_l$ frames (history).
In addition, it uses the right context of $N_r$ frames, which is future look-ahead information balancing the accuracy and latency.
Typically, non-overlapping chunk size $N_c$ (or chunk shift size) is set to 640~ms \cite{shi2021emformer,moritz20, tsunoo24_interspeech, yang24m_interspeech}, and right context frames ($N_r$) correspond to 320~ms \cite{shi2021emformer, moritz20, tsunoo24_interspeech, Li23}. 
The output is obtained only for the central $N_c$-frame chunk, leaving the left and right context.

In this paper, we consider two delays in the streaming system: internal and system-wide.
From the output of the encoder layers, the CTC module models the alignment of the tokens.
We call the internal alignment error of CTC an internal word emission delay (IWD) as in Fig.~\ref{fig:delay}.
Although the alignment is optimally modeled by minimizing IWDs, the actual output is still delayed due to block processing, as the computation of the next block only occurs after a shift of $N_c$ frames.
Therefore, we call this system-wide delay as system word emission delays (SWDs), which are also depicted in Fig.~\ref{fig:delay}.
Although the first frame of the right context already has $N_r$ frames of delay, it has to wait another $N_c$ frames to be processed for the output.
Therefore, the sum of chunk shift frames and right context frames ($N_c+N_r$) is the maximum theoretical latency, which typically corresponds to 960~ms or more.
To reduce SWDs, one can either reduce the size of chunk shift frames $N_c$ or right context frames $N_r$.
Yang {\it et al.} distill future information into small look-ahead chunks to achieve better performance, which may contribute to reducing the right context frames \cite{yang24m_interspeech}.
However, reducing the chunk shift size $N_c$ has not been well explored.

This study focuses on reducing SWDs in Transformer-based encoders \cite{vaswani17} by minimizing the chunk shift size.
Straightforwardly, 
the small chunk shift size results in frequent emission of output.
However, because it requires Transformer computation of the entire block in every $N_c$ frames, frequent computation of the Transformer increases the overall computational cost. 
Studies show that the Transformer layers have redundancy \cite{zhang2021usefulness, lee21e_interspeech,sajjad2023effect}, and thus, dropping some layers can reduce computational impact without accuracy degradation.
Early exiting techniques also contribute to reducing computational cost \cite{teerapittayanon2016branchynet,berrebbi2023avoid,yoon24_interspeech}.
It effectively stops computation midway through the layers and uses the intermediate features as output to avoid unnecessary computation of the last layers.

\begin{figure}[t]
  \centering
  \includegraphics[width=0.9\linewidth]{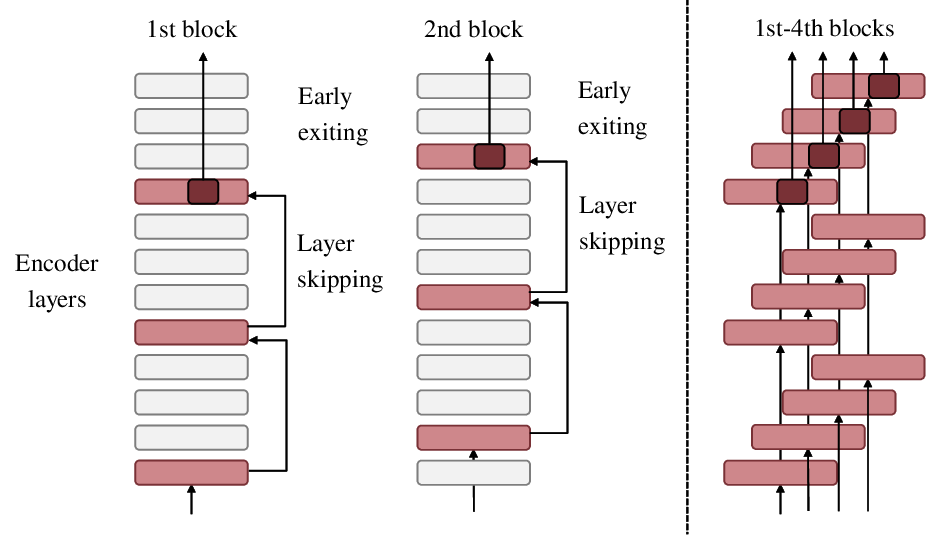}
  \caption{Circular layer skipping. Left is each block computation; right is overlapped four blocks. The computed results are cached and reused in the next block to complete ordinary full-layer computation. This is an example of a number of total layers $I=12$ and a skipping pitch $p=4$.}
  \label{fig:skipping}
\vspace{-0.3cm}
\end{figure}

Inspired by these architectural twists, we propose a new ASR encoder, namely \textit{Spiralformer}, to reduce SWDs by frequent emission with the small chunk size.
To prevent an increase in computational cost, we compute only a subset of layers at evenly spaced intervals within each block, skipping the layers in between as in Fig.~\ref{fig:skipping}.
Although the computation does not always reach to the last layer, we use the intermediate output from the last computed layer as encoded features, similarly to the early exiting technique.
In the following block computation, we alter the selection to one higher layers, as in the second block in Fig.~\ref{fig:skipping}.
To maintain its dependency on lower layer results, we combine cached intermediate output obtained in the previous block.
Thus, as the overlapping block processing goes, the layers are computed in a spiral form, which complement each other to mostly complete all the ordinary layer computation.
We conduct experiments using Librispeech \cite{panayotov15} for Englsih and Corpus of Spontaneous Japanese (CSJ) \cite{csj} for Japanese.
We compare Spiralformer with the conventional block processing of Conformer encoders \cite{gulati2020}.
In the Librispeech dataset \cite{panayotov15}, we observed that our proposed method achieved 37.5\% reduction in maximum theoretical latency, and 21.6\% reduction in empirical SWDs, compared with the baseline with similar complexity and WERs.
We also observed the similar trends in CSJ as we reduce SWDs by 7.0\% from the most competitive baseline.

\section{Block processing of encoders}
\label{sec:block}

\begin{figure}[t]
  \centering
  \includegraphics[width=0.8\linewidth]{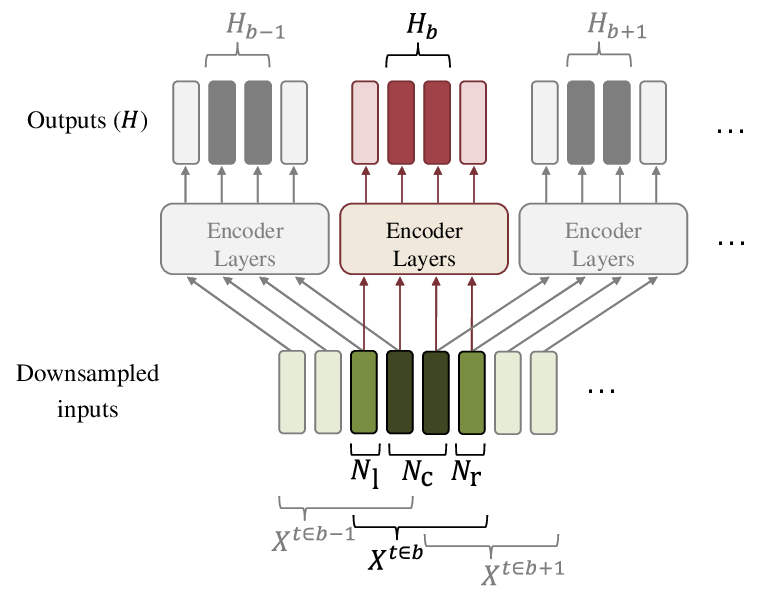}
  \caption{Blockwise processing. $N_l$ is the number of left context, $N_c$ is the number of non-overlapping chunks, and $N_r$ is the number of right context (look-ahead frames). In this example, $\{N_l, N_c, N_r\}=\{1,2,1\}$ and the encoder layers emit $N_c$-frame output $H_b$ for block $b$.}
  \label{fig:block}
\vspace{-0.3cm}
\end{figure}

Let $B$ represent the total number of blocks, and $T_b$ denote the last frame of the $b$-th block ($1\leq b \leq B$), then the speech input chunk $X^{t \in b}=\{\x_t|T_{b-1} < t \leq T_b\}$ is encoded in the $b$-th block computation.
In $X^{t \in b}$, the first $N_l$ frames are left context, following $N_c$ frames are non-overlapping central chunks, and the remaining $N_r$ frames are right context, as in Fig.~\ref{fig:block}.
Let $t$-th frame encoded feature be $\h_t$, then $N_c$-frame encoded features emitted by $b$-th block computation, which leaves $N_r$ look-ahead frames unprocessed, are described as
\begin{align}
H_b=\{\h_t|T_{b-1}+N_l<t\leq T_{b-1}+N_l+N_c\}. \label{eq:h}
\end{align}
Because the next block computation is performed only after the $N_c$ frame shift, the first frame of the right context, already having $N_r$ frames of delay, has to wait another $N_c$ frames to be processed for the output.
Therefore, it takes at most $N_c+N_r$ frames for a frame to be emitted, which is the maximum theoretical latency.
In a typical setup such as in \cite{miao2020}, i.e., $\{N_c,N_r\}=\{16,8\}$ where a frame corresponds 40~ms, the maximum theoretical latency becomes 960~ms.
With small chunk shift size $N_c$, we can expect an immediate response with a small delay.
However, it increases the frequency of block computation, i.e., $\propto$ $T_B/N_c$; thus it simply increases computational cost in total.

\section{Spiralformer}
To reduce the theoretical latency without increasing computational cost, we propose a new method that efficiently computes block processing with small $N_c$.
Inspired by dropping layers due to the redundancy of Transformer \cite{zhang2021usefulness, lee21e_interspeech,sajjad2023effect} and early exiting technique \cite{teerapittayanon2016branchynet,berrebbi2023avoid,yoon24_interspeech}, we propose an efficient method combining both, tailored for the blockwise processing.
Note that our method can be applied to any kinds of Transformer variants, such as Conformer \cite{gulati2020} or E-Branchformer \cite{kim2023branchformer}.

\subsection{Ordinary block processing}
In ordinary block processing, all the encoder layers are computed in each independent block.
Let $\Z^{(i)}_{b}$ be the intermediate output of $i$-th encode layer for block $b$, defined as
\begin{align}
    Z^{(i)}_{b} &= \enc^{(i)}(Z^{(i-1)}_b), \label{eq:singlelayer} 
\end{align}
where $\enc^{(i)}(\cdot)$ is the $i$-th encoder layer, and $Z^{(0)}_b=X^{t\in b}$.
The encoded output for the $b$-th block can be written using the last ($I$-th) layer output as $\Z^{(I)}_{b}$,
which corresponds to $N_c$ frames, as shown in Fig.~\ref{fig:block}.
For simplicity, we represent \eqref{eq:h} as 
\begin{align}
    H_b=e(Z^{(I)}_b|Z^{(I-1)}_b,Z^{(I-2)}_b,\dots,Z^{(1)}_b,X^{t\in b}) \label{eq:singleoutrewrite}
\end{align}
to explicitly indicate that the output depends on all the previous layer computations and the input.

\begin{figure}[t]
  \centering
  \includegraphics[width=0.9\linewidth]{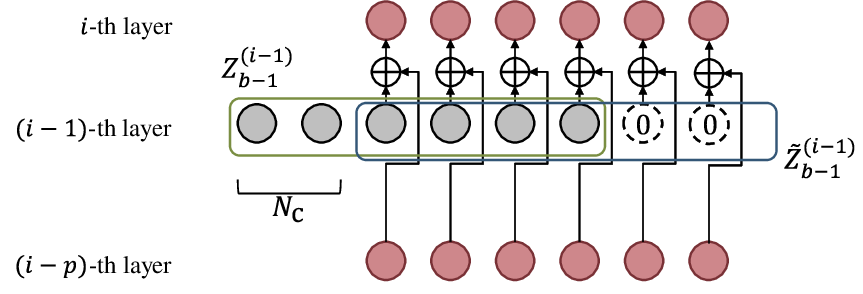}
  \caption{Layer computation with circular skipping. The $i$-th layer recurrently combines the pre-computed intermediate output of the $(i-1)$-th layer in the previous block $(b-1)$ and of the $(i-p)$-th layer in the current block ($b$).}
  \label{fig:layer}
\end{figure}

\subsection{Layer dropping}
To reduce computational cost, we drop layers and compute only a layer in every $p$ layers, where $p$ is a pitch of skipping layers.
In the next block computation, the computed layers are shifted, as in the left side of Fig.~\ref{fig:skipping}. 
Let $s$ be the shifting value defined as $s=(b-1) \bmod p$ for $b$-th block, then the set of computed layer indices are described as $\C_s=\{1+s, 1+p+s,1+2p+s,\dots\}$.
In the case of Fig.~\ref{fig:skipping}, $p=4$ and the computed layers are $\C_0=\{1,5,9\}$ for the first block, and $\C_1=\{2,6,10\}$ for the second block, and so forth.
The computation is done only in $\lfloor I/p\rfloor$ layers; thus, the complexity becomes $\lfloor I/p\rfloor/I\sim 1/p \leq1$.

\subsection{Early exiting}
We use the intermediate output from the last computed layer as the encoded features, similarly to the early exiting technique, which is highlighted in dark red in Fig.~\ref{fig:skipping}.
Therefore, each layer computation in \eqref{eq:singlelayer} and the output in \eqref{eq:singleoutrewrite} are reformulated as 
\begin{align}
    Z^{(i)}_{b} &= \begin{cases} \enc^{(i)}(X^{t\in b}) & (i\leq p) \\
    \enc^{(i)}(Z^{(i-p)}_b) & (i> p) \label{eq:spirallayer} \end{cases}\\
    H^{s}_b&=e(Z^{(I-p+s+1)}_b|Z^{i\in C_s}_b,X^{t\in b}) \nonumber\\
    &= e(Z^{(I-p+s+1)}_b|Z^{(I-{\bf 2p}+s+1)}_b,\dots,X^{t\in b}).\label{eq:skipout}
\end{align}
In the example of Fig.~\ref{fig:skipping}, $H^{0}_1=e(Z^{(9)}_1|Z^{(5)}_1,Z^{(1)}_1,X^{t\in b=1})$ for the first block, and $H^{1}_{2}=e(Z^{(10)}_{2}|Z^{(6)}_{2},Z^{(2)}_{2},X^{t\in {b=2}})$ for the second block.
Thus, only three layers are computed in each block in this case, and the complexity is reduced to 1/4.
As the block processing goes with overlapping, the computed layers shape a spiral form as shown in the right side of Fig.~\ref{fig:skipping}.

\begin{figure}[t]
  \hspace{-0.8cm}
  \includegraphics[width=1.0\linewidth]{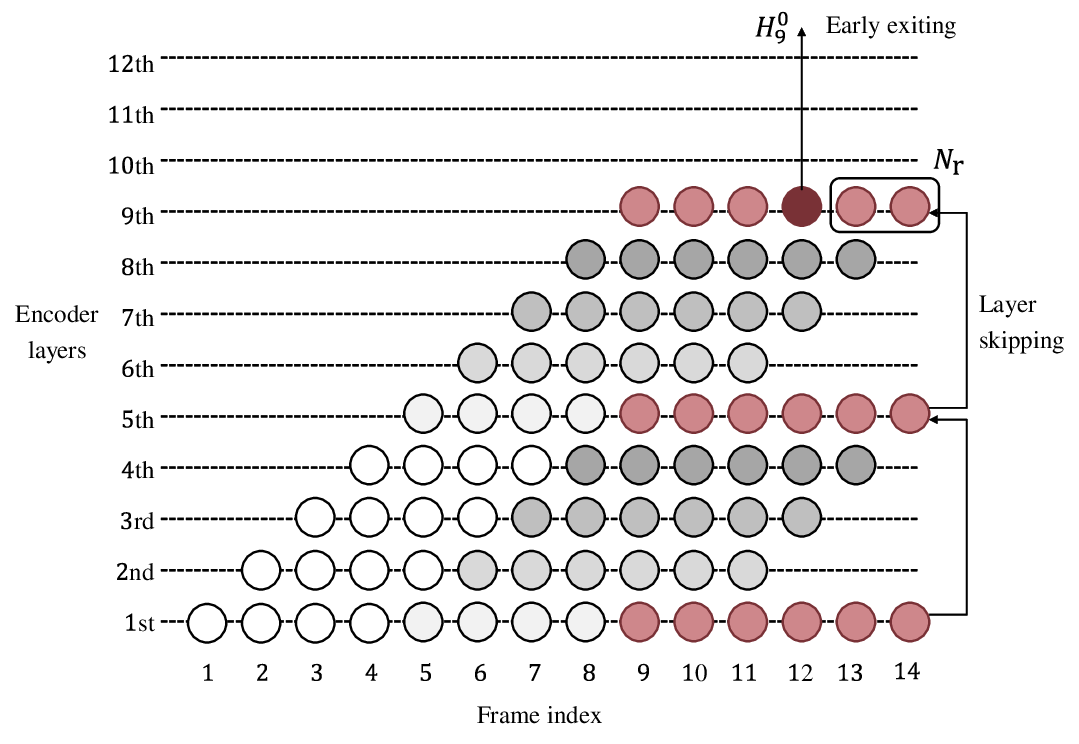}
  \caption{Example of computation accumulation. In this case, $\{N_l,N_c,N_r\}=\{3,1,2\}$ and current computing layers are $\C = \{1,5,9\}$. The layer skipping pitch is $p=4$. Because of the recurrent process of layer computation in each block, all the circled frames are considered for current block computation colored in red.}
  \label{fig:accum}
\end{figure}

\begin{table*}[t]
\caption{Encoder architecture comparison for streaming ASR trained with LibriSpeech. Spiralformer uses smaller $N_c$ to reduce maximum theoretical latency, real-time factors (RTFs), and System Word emission Delays (SWDs), simultaneously. S3 achieved similar RTF/WER with B2, reducing theoretical latency from 640~ms to 400~ms and SWD-P50 from 589~ms to 462~ms.}
\label{tab:lib960}
    \centering
    \begin{tabular}{l|c|cc|ccc|c|cc|cc}
     \hline
    Models & ID&\{$N_l$,$N_c$,$N_r$\}&$p$ & Computing & Max theoretical & Context & RTF & \multicolumn{2}{c}{WER} & \multicolumn{2}{|c}{SWD (ms)} \\
    &  & && layers &latency (ms) & length (ms) && test-clean & test-other &P50 & P90\\
    \hline\hline
         Baseline CTC \cite{tsunoo19} &B1&\{16, 16, 8\}& & 100\% & 960 &  8320 & 0.07 & 3.9 & 9.7&610 & 741\\
         &B2&\{24, 8, 8\} & & 100\% & 640 & 4800 & 0.15 & {\bf 3.4} & {\bf 8.6} & 589 & 737 \\ 
         &B3&\{28, 4, 8\}& & 100\% & 480 & 3040 & 0.32 & 3.5 & 8.8 &474 & 622\\
         &B4&\{30, 2, 8\} & & 100\% & 400 & 2160 & 0.95 & 3.5 & 9.0 & 534 & 666 \\ 
         &H2&\{28, 8, 8\} & & 50\% & 640 & 2880 & {\bf 0.05} & 4.1 & 10.7 & 515 & 666 \\ 
         &H3&\{30, 4, 8\} & & 50\% & 480 & 2080 & 0.08 & 4.1 & 10.6 & 468 & 610 \\ 
         \hline
         Spiralformer CTC&S1&\{30, 2, 8\} &4 & 25\% & 400 & 2160 &  0.11 & 4.6 & 11.4 & 398 & 472\\
         &S2&\{31, 1, 8\} & 4& 25\% & {\bf 360} & 1720 & 0.20 & 4.0 & 10.0 & {\bf 389} & {\bf 449}\\ 
         &S3&\{30, 2, 8\} & 2& 50\% & 400 & 2160 & 0.20 & 3.6& 9.1 & 462 & 522\\
         \hline
    \end{tabular}
\end{table*}

\subsection{Exploiting cached computation}
We further exploit cached computation results in the previous blocks to maintain the dependency of layer computation on lower layers.
We redefine \eqref{eq:spirallayer} to combine the previous output $Z^{(i-1)}_{b-1}$, as in Fig.~\ref{fig:layer}. 
\begin{align}
    Z^{(i)}_{b} = \left\{\begin{array} {lll}
    \enc^{(i)}(X^{t\in b} \hspace{-4mm} & + \tilde\Z^{(i-1)}_{b-1}) & (i\leq p), \\
    \enc^{(i)}(\Z^{(i-p)}_b \hspace{-4mm} & + \tilde\Z^{(i-1)}_{b-1}) &(i>p), \label{eq:layercomp}
    \end{array}
    \right.
\end{align}
where $\tilde\Z$ are pre-computed cached values from the previous blocks and used only the frames of $\Z$ corresponding to the current block with zero-padding for frames that are not pre-computed, as in Fig.~\ref{fig:layer}. 
Thus, the $i$-th layer computation depends not only on the results of the $(i-p)$-th layer as in \eqref{eq:spirallayer}, but also on the pre-computed $(i-1)$-th layer of the previous block.
Because \eqref{eq:layercomp} is calculated recurrently, this results in considering outputs of all the lower layers with longer history beyond the current block, similarly to \cite{tsunoo19, shi2021emformer}.
The output \eqref{eq:skipout} is now refomulated as 
\begin{align}
    H^{s}_b=e(Z^{(I-p+s+1)}_b|&\tilde{Z}^{(I-p+s)}_{b-1},\tilde{Z}^{(I-p+s-1)}_{b-1},\dots,Z^{(I-2p+s+1)}_b,\nonumber\\
    &\dots,X^{t\in b},X^{t\in b-1},\dots) \label{eq:spiralout}
\end{align}
Therefore, compared with \eqref{eq:skipout}, it has more dependencies on all the lower layers, and compared with \eqref{eq:singleoutrewrite}, it has similar dependencies but the previous output $\tilde{Z}$ in the skipped layers are {\it pre-computed}.
In addition, \eqref{eq:spiralout} can access to more frames than \eqref{eq:singleoutrewrite} and \eqref{eq:skipout}.
Thus, the spiral former maintains the dependencies with all layers similar to the normal block processing, while reducing the computational cost similar to the layer skipping, and increasing the receptive field similar to \cite{tsunoo19,shi2021emformer}.
Fig.~\ref{fig:accum} illustrates an example of computation of $H^0_9$ with $\{N_l,N_c,N_r\}=\{3,1,2\}$ and $\{I,i,p,s\}=\{12,9,4,0\}$.
In the 9th layer computation (colored circles), the encoder combines the 5th layer output (colored circles) and the cached 8th layer output from the previous block (the darkest gray).
This results in using previous context beyond the current analysis block, and the chain of context usage continues until it reaches the first layer; thus, all the computation results of circled frames in Fig.~\ref{fig:accum} are considered in the current computation.
Therefore, our method complementally computes all the lower layers except several newly arrived frames, unlike \eqref{eq:skipout}.

To increase coverage of the computed frames, i.e., all the circled frames in Fig.~\ref{fig:accum}, there are two choices; reducing chunk shift size $N_c$ or reducing skipping pitch $p$.
Although reducing $N_c$ increases the frequency of Transformer computation, it also increases the overlap ratio of block processing.
While reducing $p$ also increases the computational footprint, small $p$ reduces the layer dropping rate, i.e., $p=1$ is equivalent to computing an ordinary full-layer Transformer.
We will investigate the effectiveness of both options in Sec.~\ref{ssec:architecture}.

\subsection{Training Loss}
\label{ssec:loss}
In training, CTC loss $\L_{\ctc}$ \cite{graves06} is calculated using accumulated output from all the blocks, $H$ as,
\begin{align}
    \L_H &= \L_{\ctc}(Y,\CTC(H)), \\
    H &= [H^0_1,H^1_2\dots,H^{(B-1)\bmod p}_B] \label{eq:output}
\end{align}
where $\CTC(\cdot)$ is a linear function, and $Y$ is transcription.
Following \cite{yoon24_interspeech}, we also make the intermediate output of the last $p$ layers consistent by considering an additional loss for each corresponding layer.
\begin{align}
    \L^{s}_{H} &= \L_{\ctc}(Y,\CTC(H^s)) \\
    H^{s} &= [H^s_1, H^s_2,\dots,H^s_B]
\end{align}
Thus, the following combined loss is optimized in training.
\begin{align}
    \L = \L_H + \sum_{s=0}^{p-1}\L_{H}^{s} \label{eq:totalloss}
\end{align}
In inference, only the last computed layer in $\C_b$ emits the output; thus we use only $H$ in \eqref{eq:output} as the encoded features.


\section{Experiments}
\subsection{Experimental setup}
\label{ssec:setup}
To evaluate the proposed Spiralformer encoder for streaming ASR, we used the LibriSpeech \cite{panayotov15} dataset and CSJ dataset \cite{csj}.
The input acoustic features were 80-dimensional filter bank features with moving average normalization.
The encoder was a 12-block Conformer with four-head 256-unit attention layers and 2048-unit feed-forward layers unless otherwise stated.
As the baseline, we trained a 12-block Conformer with contextual blockwise processing \cite{tsunoo19}.
We trained Spiralformer initialized with the pretrained baseline Conformer.
We used a learning rate of 0.0005 without warm-up.
It was trained with the combination of CTC losses defined in \eqref{eq:totalloss}.
In addition, the external language models (LM) were used for scoring hypotheses during inference, with a fusion weight of 0.4 for Librispeech and 0.3 for CSJ.
The external LMs were two-layer LSTMs with 512 hidden units.
We applied byte-pair encoding subword tokenization \cite{sennrich16} with 5,000 token classes only for Librispeech, and used character units for CSJ.

\begin{figure}[t]
  \hspace{-0.3cm}
  \includegraphics[width=1.1\linewidth]{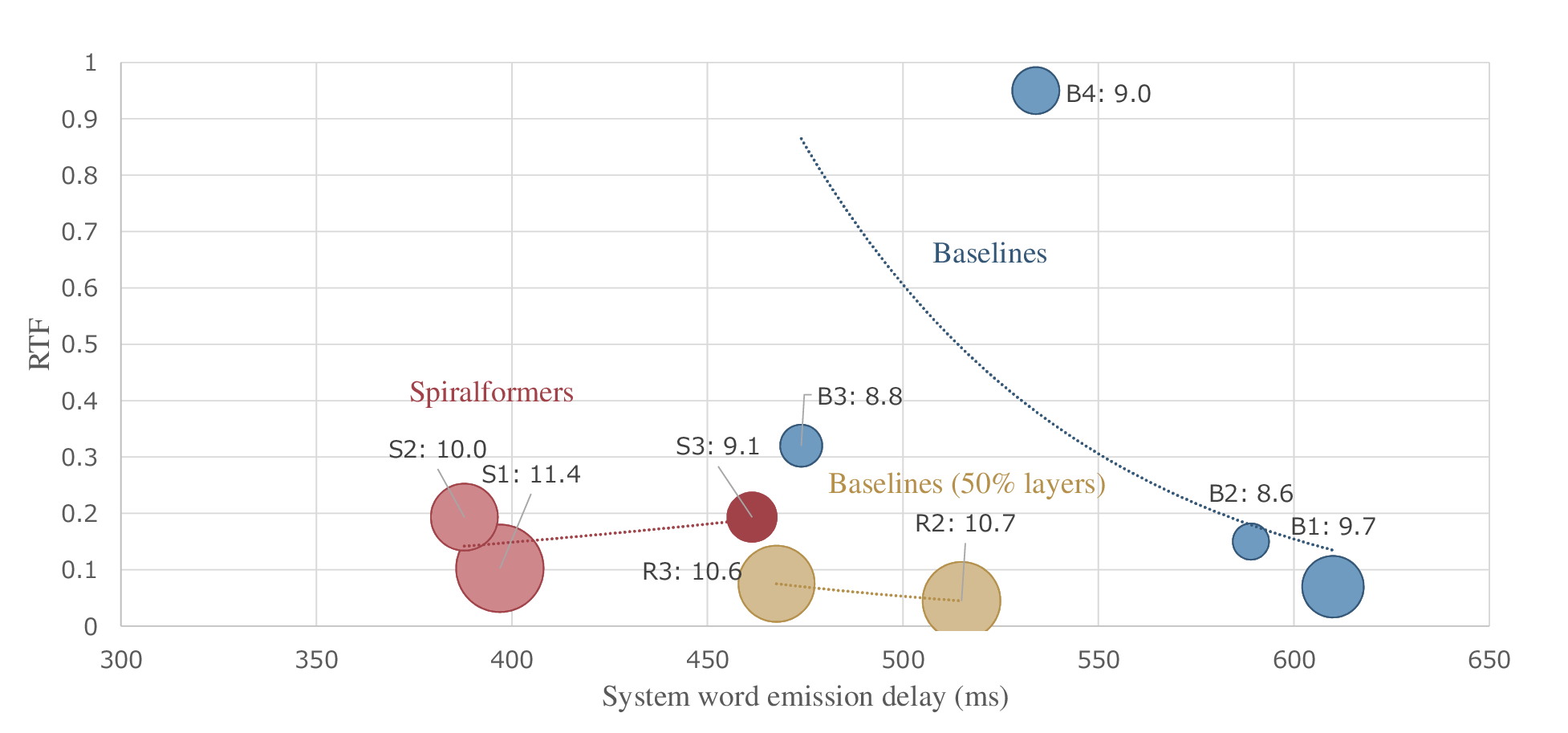}
  \caption{Visualized comparison of streaming ASR systems for Librispeech. The baseline systems are in blue, smaller baselines are in yellow, and the proposed Spiralformers are in red. The scales of circles are corresponds to WERs of test-other. Lower real-time factor (RTF) and lower system word emission delay (SWD), i.e., lower left is better.}
  \label{fig:libri_chart}
  \vspace{-0.3cm}
\end{figure}

\subsection{Encoder architecture comparison using Librispeech}
\label{ssec:architecture}
We compared various combinations of $\{N_l, N_c\}$ of both the baseline blockwise encoder and proposed Spiralformer in Table~\ref{tab:lib960}.
We fixed $N_r=8$ and varied $\{N_l, N_c\}$ because reducing the right context length was out of the scope of this study.
We measured not only WERs but also real-time factors (RTFs), which indicates computational complexity that is another factor than latency.
RTFs were measured using randomly selected 100 utterances from the test-clean set.
The proposed method was implemented in C++, and the Intel Math Kernel Library was used with an 8-core 3.60 GHz Intel i9-9900K CPU.
We maintained total analysis block size $N_l+N_c+N_r=40$ for a fair comparison.
For SWD evaluation, we measured average word emission delay averaged within each utterances as in \cite{song2023trimtail, yang24m_interspeech, shangguan21_interspeech}, which corresponds to Average Word emission Delay (AWD) in the literature.
SWD are defined as the difference between the timestamp of the word emitted completely and the timestamp of the word uttered by the speaker, as shown in Fig.~\ref{fig:delay}.
The timestamps of the uttered words were estimated using Montreal Forced Aligner \footnote{Provided in \url{https://github.com/CorentinJ/librispeech-alignments}.}, as a ground truth.
We report both the median (P50) and 90th percentile (P90) of the aforementioned 100 sampled utterances.

When we reduced the chunk shift size in the baseline model from $N_c=16$ to $N_c=2$, the Transformer computation occurred more frequently and obtained increasing RTFs as $N_c$ became smaller (B1-B4 in Table~\ref{tab:lib960}).
However, the frequent computation of the Transformer contributed to reducing WERs.
In terms of the WERs, B2 was the balanced setup for the computational frequency and its receptive field.
On the other hand, our proposed method successfully reduced computational costs (RTFs).
When $N_c=2$ and $p=4$ (S1), the RTF was significantly reduced from 0.95 of B4 to 0.11, with the same maximum theoretical latency of 400~ms.
However, dropping many layers, resulting in a 75\% reduction of computation, lost the model capacity, and we observed significant degradation in WERs from 3.5/9.0 in B4 to 4.6/11.4.

To increase model capacity, we increase the coverage of computed frames in Fig.~\ref{fig:accum} by adopting either smaller $N_c$ or smaller skipping pitch $p$.
With the half chunk size $N_c=1$ (S2), we observed accuracy improvement at the cost of computational cost, which increased by almost double because the half chunk size increased the computation frequency twofold.
Reducing the skipping pitch to $p=2$ (S3) also doubled the computational cost.
However, WER reduction was larger, from 4.6/11.4 to 3.6/9.1 (S1 vs S3), which was comparable to B4 having a higher RTF.
This also indicated that smaller skipping pitch $p$ was more efficient to improve performance than smaller $N_c$.
We concluded that $N_c=2$ and $p=2$ (S3) were the best balanced in this dataset for Spiralformer.
We also prepared baseline models with the half size, i.e., six layers, of B2 and B3 as H2 and H3 to compare with 50\%-layer computation of S3.
However, the WERs were significantly higher than S3.
These results also indicate the efficiency of the proposed architecture.

For SWD, Spiralformer (S1) achieved a smaller delay than the baseline (B1), i.e., 398~ms against 610~ms at P50, owing to the small chunk shift size $N_c$.
In some moldes such as B4, the SWD (534~ms at P50) exceeded the maximum theoretical latency (400~ms).
We suppose that, to optimize recognition accuracy, the CTC decision was deferred, which drifted emission timing behind, i.e., IWD $> 0$ in Fig.~\ref{fig:delay}.
S3 achieved a significant reduction in SWD, compared with a similar range of computational cost and WERs of B2 (462~ms against 589~ms at P50).
The latency reduction resulted in 21.6\% at P50.

Because there are three aspect to be compared, i.e.,  WERs, RTFs, and SWDs, we also graphically show the same results of Table.~\ref{tab:lib960} in Fig.~\ref{fig:libri_chart}.
The horizontal axis is SWD, the vertical axis is RTF, and the scale of each circle is the WER of the test-other set.
In all aspects, the lower the better; thus smaller lower left circles represent better systems.
Although half baselines (H2 and H3) have smaller SWDs/RTFs, the WERs are significantly higher than other models.
On the other hand, our proposed Spiralformer S3 is located in lower left than the baseline models, indicating that it improves overall system performance.

\begin{table}[t]
\caption{Emission delay on the Librispeech test-clean set. First Word emission Delay (FWD), Last Word emission Delay (LWD), and averaged System Word emission Delay (SWD) are evaluated at 50\%/90\% percentiles. Spiralformer generally achieves lower FWD and SWD.}  
\label{tab:latency}
    \centering
    \begin{tabular}{l|c|cccccc}
        \hline
         Models & ID&\multicolumn{2}{c}{FWD (ms)} & \multicolumn{2}{c}{LWD (ms)} & \multicolumn{2}{c}{SWD (ms)} \\
         & & P50 & P90 & P50 & P90 & P50 & P90 \\
         \hline\hline
         Baseline CTC \cite{tsunoo19} &B1& 980 & 1200 & 430 &520 & 610 & 741 \\
         &B2& 990 & 1250 & 430 & 520 & 589 & 737 \\
         &B3& 990 & 1210 & 430 & 520 & 474 & 622 \\
         &B4& 990 & 1210 & 430 & 520 & 534 & 666 \\
         &H2& 980 & 1240 & 430 & 520 & 515 & 666\\
         &H3& 1000 & 1240 & 430 & 520 & 468 & 610 \\
         \hline
         Spiralformer CTC&S1 &440 & 510 & 430 & 520 & 398 & 472 \\
         &S2& {\bf 420} & {\bf 500} & 430 & 520  & {\bf 389} & {\bf 449} \\
         &S3& 510 & 590 & 430 & 520 & 462 & 522 \\
         \hline
    \end{tabular}
\end{table}

As in \cite{song2023trimtail, yang24m_interspeech, shangguan21_interspeech}, we also measured First Word emission Delay (FWD), Last Word emission Delays (LWD) in addition to SWD, because FWD and LWD are especially important aspect for the user experience in real-world streaming ASR applications as discussed in \cite{song2023trimtail}.
The results are listed in Table.~\ref{tab:latency}.
All the LWDs look the same because we used the same right context $N_r$, and the last frames were consumed at once in the end.
We observed significant reduction in FWDs with our proposed Spiralformer (S1--S3), contributing to the better usability.

\begin{table*}[t]
\caption{Performance comparison for streaming ASR trained with CSJ. }
\label{tab:csj}
    \centering
    \begin{tabular}{l|c|cc|cc|c|ccc|cc}
     \hline
    Models & ID&\{$N_l$,$N_c$,$N_r$\}&$p$ & Computing & Max theoretical & RTF & \multicolumn{3}{c}{CER} & \multicolumn{2}{|c}{SWD (ms)} \\
    &  & && layers &latency (ms) && eval1 & eval2 & eval3 &P50 & P90\\
    \hline\hline
         Baseline CTC \cite{tsunoo19} &B1&\{16, 16, 8\}& & 100\% & 960 & {\bf 0.05} & 5.7 & 4.3&  4.8&585 & 590 \\
         &B2&\{24, 8, 8\} & & 100\% & 640  & 0.09 & 5.7 & 4.1 &  4.8 & 442 &  454\\ 
         &B3&\{28, 4, 8\}& & 100\% & 480  & 0.17 & {\bf 5.4} & {\bf 3.9} & {\bf 4.5}  &385  & 396\\
         &B4&\{30, 2, 8\} & & 100\% & 400  & 0.39 & 5.5 & 4.0 & {\bf 4.5} & 359 & 369  \\ 
         &H2&\{24, 8, 8\} & & 50\% & 640  & {\bf 0.05} & 6.6 & 4.7 & 5.3 & 422 & 433  \\ 
         &H3&\{28, 4, 8\} & & 50\% & 480  & 0.09 & 6.4& 4.4 & 5.2 &  354 & 366  \\ 
         \hline
         Spiralformer CTC&S1&\{30, 2, 8\} &4 & 25\% & 400 & 0.11 & 6.8& 5.1& 5.8 & 367 & 374\\
         &S2&\{31, 1, 8\} & 4& 25\% & {\bf 360}  & 0.20 & 6.0&4.4&5.2& {\bf 345} & {\bf 350} \\ 
         &S3&\{30, 2, 8\} & 2& 50\% & 400 & 0.20 &  {\bf 5.4}& 4.2 & 4.8 & 358 & 366\\
         \hline
    \end{tabular}
\end{table*}

\begin{table*}[t]
\caption{Ablation study on training strategy for streaming ASR models trained with LibriSpeech. Each Spiralformer was trained from scratch or finetuned from the baseline model. We evaluated WERs as well as internal word emission Delay (IWD) of CTC emissions, and System Word emission Delay (SWD).}
\label{tab:training}
    \centering
    \begin{tabular}{l|c|ccc|cc|cc|cc}
     \hline
    Models & ID&\{$N_l$,$N_c$,$N_r$\}& $p$ & Computing & \multicolumn{2}{|c}{WER}&\multicolumn{2}{|c}{IWD (ms)} & \multicolumn{2}{|c}{SWD (ms)} \\
    &  & && layers &test-clean & test-other& P50 & P90 & P50 & P90  \\
    \hline\hline
         Spiralformer CTC&S1&\{30, 2, 8\} &4 & 25\% &  4.6 & 11.4 & {\bf 21} & {\bf 54} & {\bf 398} & {\bf 472}\\ 
         \multicolumn{1}{r|}{\it from scratch} &S1FS&\{30, 2, 8\} &4 & 25\% &  {\bf 4.4} & {\bf 11.3}  & 132 & 155 & 513 & 572\\ 
         \hline
    \end{tabular}
\end{table*}

\subsection{Evaluation results on CSJ}
We conducted the same experiments in a different language using CSJ dataset.
For the emission delay evaluation, we used official annotation of timestamps as a ground truth.
Note that the timestamps were at the morpheme-level, while the ASR models are trained with character units.

\begin{figure}[t]
  \centering
  \hspace{-0.3cm}
  \includegraphics[width=1.1\linewidth]{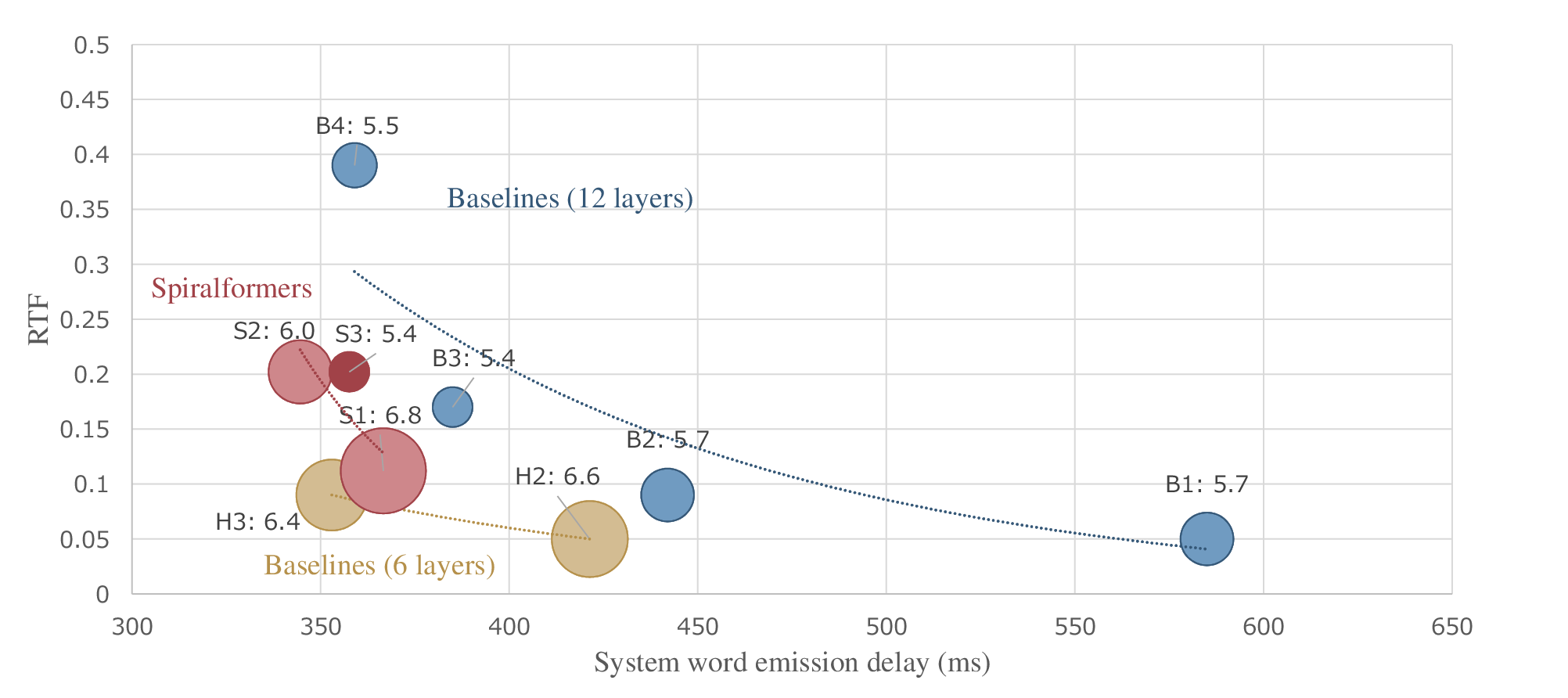}
  \caption{Visualized comparison of streaming ASR systems for CSJ.}
  \label{fig:csj_chart}
  \vspace{-0.3cm}
\end{figure}

The results are listed in Table~\ref{tab:csj} and its visualization is in Fig.~\ref{fig:csj_chart}.
The results show the similar trends as in Table~\ref{tab:lib960} of Librispeech.
As the baseline, we compared the various parameter setups and the more transformer computation resulted in lower character error rates (CERs) with higher computational complexity (RTFs).
When we halved the number of layers to six (H2 and H3), which is comparable to 50\% skipping of S3, the CERs degraded significantly.
The proposed S3 achieved comparable CERs, and RTF to the baseline B2 or B3, similarly to Table ~\ref{tab:lib960}.
S3 also reduced SWDs compared with the competitive baseline of B2 or B3.
The reduction at P50 was 19.0\% from B2 and 7.0\% from B3.

\subsection{Ablation study on training strategy}
\label{ssec:training}
Lastly, we conducted an ablation study for the training strategy.
Using Librispeech, we compared both training Spiralformer from scratch and finetuning the pretrained Conformer with contextual block processing as in Sec.~\ref{ssec:architecture}.
When we trained from scratch, we used a learning rate of 0.0025 with Noam learning rate decay.
We analyzed internal word emission delay (IWD) in Fig.~\ref{fig:delay} using the same subset for RTF computation in Sec.~\ref{ssec:architecture}.
We computed CTC emission probabilities over the accumulated encoder output ($H$ in \eqref{eq:output}) and compared argmax (greedy decoding) to the forced alignment ground truth.
This evaluation is similar to the scenarios in \cite{yu2021fastemit,song2023trimtail}.
The results are listed in Table~\ref{tab:training}.

Finetuned Spiralformer (S1) successfully aligned with the ground truth, with only a 21~ms difference at IWD-P50.
On the other hand, CTC emission timestamps in Spiralformer trained from scratch (S1FS) shifted 132~ms behind the ground truth at IWD-P50.
This may be because, to optimize recognition accuracy, the CTC decision was deferred, and it was possible due to the context passing mechanism in \eqref{eq:layercomp}.
This drift of S1FS also affected SWDs compared with S1 (513~ms vs 398~ms at P50).
Although training from scratch (S1FS) could achieve lower WERs than the finetuned model (S1), the difference was not significant and finetuning strategy had an advantage in emission timing.

\section{Conclusion}
We investigated reducing latency in the encoder for block processing with small chunk shift size.
To mitigate increasing computational cost with frequent Transformer computation, we proposed Spiralformer, a new block processing encoder combining layer dropping and early exiting techniques.
We skipped layers with the cyclic layer selection, and the selection was shifted over the block processing using cached pre-computed results.
Thus, after the blocks were processed with overlap, the computed layers complemented each other to complete ordinary Transformer computation.
Experimentally, we compared our proposed method with the conventional block processing of Conformer encoders.
In the Librispeech dataset, we observed that our proposed method was processed with smaller maximum theoretical latencies and smaller empirical latencies, compared with the baseline with the similar WERs and computational footprint.
We observed that our proposed method achieved 37.5\% reduction in maximum theoretical latency, and 21.6\% reduction in empirical SWD.
Similar trends were observed in CSJ dataset, as we reduce SWD by 7.0\% from the most competitive baseline.

Future work includes regularized training to prevent emission drifting of the Spiralformer CTC output.
Investigation on decoders that exploit this efficient encoder architecture will also be investigated.

\newpage

\bibliographystyle{IEEEtran}
\bibliography{mybib}

\begin{thebibliography}{10}
\providecommand{\url}[1]{#1}
\csname url@samestyle\endcsname
\providecommand{\newblock}{\relax}
\providecommand{\bibinfo}[2]{#2}
\providecommand{\BIBentrySTDinterwordspacing}{\spaceskip=0pt\relax}
\providecommand{\BIBentryALTinterwordstretchfactor}{4}
\providecommand{\BIBentryALTinterwordspacing}{\spaceskip=\fontdimen2\font plus
\BIBentryALTinterwordstretchfactor\fontdimen3\font minus
  \fontdimen4\font\relax}
\providecommand{\BIBforeignlanguage}[2]{{%
\expandafter\ifx\csname l@#1\endcsname\relax
\typeout{** WARNING: IEEEtran.bst: No hyphenation pattern has been}%
\typeout{** loaded for the language `#1'. Using the pattern for}%
\typeout{** the default language instead.}%
\else
\language=\csname l@#1\endcsname
\fi
#2}}
\providecommand{\BIBdecl}{\relax}
\BIBdecl

\bibitem{povey18}
D.~Povey, H.~Hadian, P.~Ghahremani, K.~Li, and S.~Khudanpur, ``A
  time-restricted self-attention layer for {ASR},'' in \emph{Proc. of ICASSP},
  2018, pp. 5874--5878.

\bibitem{dong19}
L.~Dong, F.~Wang, and B.~Xu, ``Self-attention aligner: A latency-control
  end-to-end model for {ASR} using self-attention network and chunk-hopping,''
  in \emph{Proc. of ICASSP}, 2019, pp. 5656--5660.

\bibitem{tsunoo19}
E.~Tsunoo, Y.~Kashiwagi, T.~Kumakura, and S.~Watanabe, ``Transformer {ASR} with
  contextual block processing,'' in \emph{Proc. of ASRU Workshop}, 2019, pp.
  427--433.

\bibitem{miao2020}
H.~Miao, G.~Cheng, Z.~Pengyuan, and Y.~Yan, ``Transformer-based online
  {CTC}/attention end-to-end speech recognition architecture,'' in \emph{Proc.
  of ICASSP}, 2020, pp. 6084--6088.

\bibitem{shi2021emformer}
Y.~Shi, Y.~Wang, C.~Wu, C.-F. Yeh, J.~Chan, F.~Zhang, D.~Le, and M.~Seltzer,
  ``Emformer: Efficient memory transformer based acoustic model for low latency
  streaming speech recognition,'' in \emph{Proc. of ICASSP}, 2021, pp.
  6783--6787.

\bibitem{graves06}
A.~Graves, S.~Fern{\'a}ndez, F.~Gomez, and J.~Schmidhuber, ``Connectionist
  temporal classification: labelling unsegmented sequence data with recurrent
  neural networks,'' in \emph{Proc. of 23rd International Conference on Machine
  Learning}, 2006, pp. 369--376.

\bibitem{miao15}
Y.~Miao, M.~Gowayyed, and F.~Metze, ``{EESEN}: End-to-end speech recognition
  using deep {RNN} models and {WFST}-based decoding,'' in \emph{Proc. of ASRU
  Workshop}, 2015, pp. 167--174.

\bibitem{amodei16}
D.~Amodei \emph{et~al.}, ``Deep {Speech} 2: End-to-end speech recognition in
  {English} and {Mandarin},'' in \emph{Proc. of 33rd International Conference
  on Machine Learning}, vol.~48, 2016, pp. 173--182.

\bibitem{arora2023semi}
S.~Arora, G.~Saon, S.~Watanabe, and B.~Kingsbury, ``Semi-autoregressive
  streaming {ASR} with label context,'' \emph{arXiv preprint arXiv:2309.10926},
  2023.

\bibitem{graves13rnnt}
A.~Graves, A.-R. Mohamed, and G.~Hinton, ``Speech recognition with deep
  recurrent neural networks,'' in \emph{Proc. of ICASSP}, 2013, pp. 6645--6649.

\bibitem{rao17}
K.~Rao, H.~Sak, and R.~Prabhavalkar, ``Exploring architectures, data and units
  for streaming end-to-end speech recognition with {RNN}-transducer,'' in
  \emph{Proc. of ASRU Workshop}, 2017, pp. 193--199.

\bibitem{zhang2020transformer}
Q.~Zhang, H.~Lu, H.~Sak, A.~Tripathi, E.~McDermott, S.~Koo, and S.~Kumar,
  ``Transformer transducer: A streamable speech recognition model with
  transformer encoders and {RNN-T} loss,'' in \emph{Proc. of ICASSP}, 2020, pp.
  7829--7833.

\bibitem{li2020towards}
B.~Li, S.-y. Chang, T.~N. Sainath, R.~Pang, Y.~He, T.~Strohman, and Y.~Wu,
  ``Towards fast and accurate streaming end-to-end {ASR},'' in \emph{Proc.
  ICASSP}, 2020, pp. 6069--6073.

\bibitem{sainath2020emitting}
T.~N. Sainath, R.~Pang, D.~Rybach, B.~Garc{\'\i}a, and T.~Strohman, ``Emitting
  word timings with end-to-end models.'' in \emph{Proc. of Interspeech}, 2020,
  pp. 3615--3619.

\bibitem{yu2021fastemit}
J.~Yu, C.-C. Chiu, B.~Li, S.-y. Chang, T.~N. Sainath, Y.~He, A.~Narayanan,
  W.~Han, A.~Gulati, Y.~Wu \emph{et~al.}, ``{FastEmit}: Low-latency streaming
  {ASR} with sequence-level emission regularization,'' in \emph{Proc. of
  ICASSP}, 2021, pp. 6004--6008.

\bibitem{song2023trimtail}
X.~Song, D.~Wu, Z.~Wu, B.~Zhang, Y.~Zhang, Z.~Peng, W.~Li, F.~Pan, and C.~Zhu,
  ``{TrimTail}: {Low-latency} streaming {ASR} with simple but effective
  spectrogram-level length penalty,'' in \emph{Proc. of ICASSP}, 2023, pp.
  1--5.

\bibitem{moritz20}
N.~Moritz, T.~Hori, and J.~L. Roux, ``Streaming automatic speech recognition
  with the transformer model,'' in \emph{Proc. of ICASSP}, 2020, pp.
  6074--6078.

\bibitem{tsunoo24_interspeech}
E.~Tsunoo, H.~Futami, Y.~Kashiwagi, S.~Arora, and S.~Watanabe, ``Decoder-only
  architecture for streaming end-to-end speech recognition,'' in \emph{Proc. of
  Interspeech}, 2024, pp. 4463--4467.

\bibitem{yang24m_interspeech}
Y.~Yang, G.~Ma, Y.~Li, B.~Du, H.~Zhu, and L.~Ruan, ``Learning from back chunks:
  Acquiring more future knowledge for streaming {ASR} models via self
  distillation,'' in \emph{Proc. of Interspeech}, 2024, pp. 4458--4462.

\bibitem{Li23}
M.~Li, S.~Zhang, C.~Zorilă, and R.~Doddipatla, ``Transformer-based streaming
  {ASR} with cumulative attention,'' in \emph{Proc. of ICASSP}, 2022, pp.
  8272--8276.

\bibitem{vaswani17}
A.~Vaswani, N.~Shazeer, N.~Parmar, J.~Uszkoreit, L.~Jones, A.~N. Gomez,
  {\L}.~Kaiser, and I.~Polosukhin, ``Attention is all you need,'' in
  \emph{Proc. of NeurIPS}, 2017, pp. 5998--6008.

\bibitem{zhang2021usefulness}
S.~Zhang, E.~Loweimi, P.~Bell, and S.~Renals, ``On the usefulness of
  self-attention for automatic speech recognition with transformers,'' in
  \emph{2021 IEEE Spoken Language Technology Workshop (SLT)}, 2021, pp. 89--96.

\bibitem{lee21e_interspeech}
J.~Lee, J.~Kang, and S.~Watanabe, ``Layer pruning on demand with intermediate
  {CTC},'' in \emph{Proc. of Interspeech}, 2021, pp. 3745--3749.

\bibitem{sajjad2023effect}
H.~Sajjad, F.~Dalvi, N.~Durrani, and P.~Nakov, ``On the effect of dropping
  layers of pre-trained transformer models,'' \emph{Computer Speech \&
  Language}, vol.~77, p. 101429, 2023.

\bibitem{teerapittayanon2016branchynet}
S.~Teerapittayanon, B.~McDanel, and H.-T. Kung, ``Branchynet: {Fast} inference
  via early exiting from deep neural networks,'' in \emph{2016 23rd
  international conference on pattern recognition (ICPR)}, 2016, pp.
  2464--2469.

\bibitem{berrebbi2023avoid}
D.~Berrebbi, B.~Yan, and S.~Watanabe, ``Avoid overthinking in self-supervised
  models for speech recognition,'' in \emph{Proc. of ICASSP}, 2023, pp. 1--5.

\bibitem{yoon24_interspeech}
J.~W. Yoon, B.~J. Woo, and N.~S. Kim, ``{HuBERT-EE}: {Early} exiting {HuBERT}
  for efficient speech recognition,'' in \emph{Proc. of Interspeech}, 2024, pp.
  2400--2404.

\bibitem{panayotov15}
V.~Panayotov, G.~Chen, D.~Povey, and S.~Khudanpur, ``{LibriSpeech}: an {ASR}
  corpus based on public domain audio books,'' in \emph{Proc. of ICASSP}, 2015,
  pp. 5206--5210.

\bibitem{csj}
K.~Maekawa, H.~Koiso, S.~Furui, and H.~Isahara, ``Spontaneous speech corpus of
  {Japanese},'' in \emph{Proc. of the International Conference on Language
  Resources and Evaluation (LREC)}, 2000, pp. 947--9520.

\bibitem{gulati2020}
A.~Gulati, J.~Qin, C.-C. Chiu, N.~Parmar, Y.~Zhang, J.~Yu, W.~Han, S.~Wang,
  Z.~Zhang, Y.~Wu \emph{et~al.}, ``Conformer: Convolution-augmented transformer
  for speech recognition,'' in \emph{Proc. of Interspeech}, 2020, pp.
  5036--5040.

\bibitem{kim2023branchformer}
K.~Kim, F.~Wu, Y.~Peng, J.~Pan, P.~Sridhar, K.~J. Han, and S.~Watanabe,
  ``E-branchformer: Branchformer with enhanced merging for speech
  recognition,'' in \emph{2022 IEEE Spoken Language Technology Workshop (SLT)},
  2023, pp. 84--91.

\bibitem{sennrich16}
R.~Sennrich, B.~Haddow, and A.~Birch, ``Neural machine translation of rare
  words with subword units,'' in \emph{Proc. of the Association for
  Computational Linguistics}, vol.~1, 2016, pp. 1715--1725.

\bibitem{shangguan21_interspeech}
Y.~Shangguan, R.~Prabhavalkar, H.~Su, J.~Mahadeokar, Y.~Shi, J.~Zhou, C.~Wu,
  D.~Le, O.~Kalinli, C.~Fuegen, and M.~L. Seltzer, ``Dissecting user-perceived
  latency of on-device {E2E} speech recognition,'' in \emph{Proc. of
  Interspeech}, 2021, pp. 4553--4557.

\end{thebibliography}

\end{document}